\documentclass[psfig]{mn2e}
\usepackage{amssymb}
\usepackage{psfig,epsf}

\begin{document}

\title[The Universal Rotation Curve of Spiral Galaxies]{The Universal
Rotation Curve of Spiral Galaxies}
\author[A.A. Kirillov and D. Turaev]{A.A. Kirillov $^{1}$\thanks{%
E-mail:ka98@mail.ru} and D. Turaev$^{2}$\thanks{%
E-mail:turaev@math.bgu.ac.il} \\
$^{1}$ Institute for Applied Mathematics and Cybernetics, 10 Uljanova Str.,
Nizhny Novgorod, 603005, Russia\\
$^{2}$ Ben-Gurion University of the Negev, P.O.B. 653, Beer-Sheva 84105,
Israel}
\date{Accepted 2006 June 12. Received 2006 June 6; in original form 2006 April 24}
\maketitle

\begin{abstract}
The observed strong dark-to-luminous matter coupling is described by a bias
relation between visible and dark matter sources. We discuss the bias which
emerges in the case where the topological structure of the Universe at very
large distances does not match properly that of the Friedman space. With the
use of such "topological" bias, we construct the Universal Rotation Curve
(URC) for spirals which occurs to be in a striking agreement with the
empirically known URC. We also show that the topological bias explains the
origin of the Tully-Fisher relation ($L\sim V^{4}$) and predicts peculiar
oscillations in the URC with a characteristic length $\sim \sqrt{L}$.
\end{abstract}


\section{Introduction}

It has been long known (Persic, Salucci \& Stel 1996, PSS
hereafter) that the shape of rotation curves of spirals is rigidly
determined by a single global parameter, e.g., luminosity or the
number of baryons in a galaxy. This feature was stressed in PSS by
an empirical construction of a universal rotation curve (URC)
which describes quite well the rotation velocity at any radius and
for any galaxy as a function of, say, the galaxy luminosity only.
It follows that the distribution of the Dark Matter (DM) in
galaxies carries a universal character as well, and is a function
of the luminous mass. Note that the standard Cold DM (CDM) models
fail to explain this strong dark-to-luminous matter coupling, by
an obvious reason: in any model where DM is built from
hypothetical non-baryonic particles (e.g., CDM, worm DM, or self-
interacting DM) the number of the DM particles in the halo of a
galaxy is, essentially, a free parameter and relating it to the
number of baryons in the galaxy requires some very strong
nonlinearity. Moreover, it is well established that the DM density
in galaxies shows an inner core, i.e. a central constant density
region (e.g., see Gentile et al., 2004 and references therein;
Weldrake, de Blok \& Walter 2003; de Blok \& Bosma 2002; for
spirals and Gerhard et al., 2001; Borriello, Salucci \& Danese
2003 for ellipticals), which is in a clear conflict with the
predictions of $\Lambda $ --CDM models yielding NFW-type profiles
with a cusp (e.g. $\rho _{DM}\sim 1/r $) in the central region of
a galaxy (Navarro, Frenk \& White 1996).

The strong coupling between DM halos and baryons (see also Donato, Gentile
\& Salucci 2004) definitely requires some new physics. The coupling can be
described by a rigid relation between the sources of dark, $\rho _{DM}$, and
visible, $\rho _{L}$, matter, the so-called bias relation (Kirillov 2006).
In the linear case the most general form of the bias relation is
\begin{equation}
\rho _{DM}\left( x\right) =\int b\left( x,x^{\prime }\right) \rho _{L}\left(
x^{\prime }\right) dV^{\prime }.  \label{br1}
\end{equation}
The homogeneity assumption $b\left( x,x^{\prime }\right) =b\left(
x-x^{\prime }\right) $ allows one to fix empirically the bias
operator $b_{emp}$. Indeed, in this case the Fourier transform of
the bias relation (\ref{br1}) gives
\begin{equation}
\rho _{DM}\left( t,k\right) =b\left( t,k\right) \ \rho _{L}\left( t,k\right)
\label{br2}
\end{equation}
where we added a dependence on time to account for the cosmic evolution. The
empirical bias function $b_{emp}=$ $\rho _{DM}\left( t,k\right) /\rho
_{L}\left( t,k\right) $, in virtue merely of its definition, will perfectly
describe DM effects at very large scales (i.e. in the region of linear
perturbations).

The present Universe is not quite homogeneous though, e.g. it is
not uniform at galaxy scales. Still we would expect relation
(\ref{br2}) to hold in the geometrical optics limit (i.e. for
rather short wavelengths as compared to the Hubble scale, or to a
cluster scale when a single galaxy is considered). Parameters of
the bias function may then vary for different spatial regions,
i.e. $b_{emp}$ may include an additional slow dependence on the
location in space: $b=b_{emp}\left( t,k,x\right) $. In order to
fit observations, any theoretical source of DM should reproduce
properties of the bias function $b_{emp}$ in detail.

In the linear gravity, the bias relation (\ref{br2}) can be interpreted as a
modification of the Newton law:
\begin{equation}
\frac{1}{r^{2}}\rightarrow \frac{2}{\pi }\int\limits_{0}^{\infty }(1+b\left(
k\right) )\frac{\sin \left( kr\right) -kr\cos (kr)}{kr^{2}}dk.  \label{br3}
\end{equation}
The asymptotically flat rotation curves in galaxies require that the
correction to the Newton's potential should be logarithmic, i.e. the
gravitational acceleration should switch from $r^{-2}$ to $r^{-1}$. This,
according to (\ref{br3}), implies $b(k)\sim k^{-1}$, or
\begin{equation}
b(x-x^{\prime })\sim |x-x^{\prime }|^{-2}  \label{br4}
\end{equation}
at galaxy scales. \ In fact, observations suggest the same
behavior of $b(x-x^{\prime })$ for much larger scales (Kirillov
2006). Indeed, the distribution of the luminous mass shows
characteristically fractal behavior: the mass $M_{L}(r)$ within
the ball of radius $r$ grows, essentially, as  $\sim r^{D}$ with
$D\simeq 2$ on distances up to at least $200$Mpc (Pietronero 1987;
Ruffini, Song \& Taraglio 1988, Labini, Montuori \& Pietronero
1998). The bias function (\ref{br4}) leads then to $M_{DM}(r)\sim
r^{3}$, i.e. to a homogeneous distribution of the DM and, hence,
of the total mass. Thus, bias (\ref{br4}) is just one that
reconciles the two seemingly contradictory observational facts:
the fractal distribution of baryons with the dimension $D\simeq 2$
and the large-scale homogeneity of the metric.

A theoretical scheme capable of explaining the origin of such bias was
proposed in (Kirillov 1999, Kirillov \& Turaev 2002). It was shown there
that processes involving topology changes during the quantum stage of the
evolution of the Universe unavoidably (and, in fact, model-independently)
lead to a scale-dependent renormalization of the constant of gravity $G$,
and this effect can be imitated by the emergence of DM, whose distribution
is linearly related to the distribution of actual matter. Importantly,
assuming the thermal equilibrium during the quantum stage predicts in almost
unique way a very specific form of the bias function (Kirillov \& Turaev
2002; Kirillov 2003)
\begin{equation}
b\left( k\right) =\frac{\mu }{\sqrt{\kappa ^{2}+k^{2}}}\ \ \
\mbox{for }\ \ k<\mu .  \label{b3}
\end{equation}%
where $\mu $ $\sim T_{Pl}a\left( t_{Pl}\right) /a\left( t\right) $
has the meaning of the primordial temperature at which the
topology has been tempered and $\kappa \sim ma\left( t_{Pl}\right)
/a\left( t\right) $ is the mass of primordial scalar particles if
they do exist. This means that at scales $k\lesssim \kappa $ the
bias becomes the constant $b\left( k\right) =\mu /\kappa $ and the
standard Newton's law restores. However at galaxy scales it should
be $k\sim \mu \gg \kappa $ and in what follows we take $\kappa =0$
in (\ref{b3}) \footnote{E.g., inflationary scenarios suggest
$m\sim 10^{-5}m_{Pl}$ and, therefore, $\mu /\kappa \sim 10^{5}$.
However so far scalar particles have not been observed which makes
us think that scalar fields are not more but phenomenological
objects.}. Thus in the coordinate representation the bias takes
the form
\begin{eqnarray}
b\left( \vec{r},t\right) &=&\frac{1}{2\pi ^{2}}\int\limits_{0}^{\mu }\left(
b\left( k\right) k^{3}\right) \frac{\sin \left( kr\right) }{kr}\frac{dk}{k}=
\label{b4} \\
&=&\frac{\mu }{2\pi ^{2}r^{2}}\left( 1-\cos \left( \mu r\right) \right) .
\nonumber
\end{eqnarray}%
Bias (\ref{b3}) is of form (\ref{br4}), so it predicts the logarithmic
correction to the Newton's potential for a point source: $\delta \phi \sim
\frac{1}{R_{0}}\ln r$ \ at $r>R_{0}$ where $R_{0}=\pi /\left( 2\mu \right) $
(see for details Kirillov \& Turaev 2002; Kirillov 2006). Thus, the
parameter $R_{0}$ plays the role of the scale at which DM starts to show up,
so in galaxies it has to be estimated as a few Kpc. As bias (\ref{b3}) has a
thermodynamical origin, there have to be certain fluctuations in the value
of $R_{0}$ (this effect is analyzed in the next Section).

In the present Letter we demonstrate that bias (\ref{b3}), (\ref{b4}) gives
a very good agreement with the empirical URC constructed in PSS and,
therefore, it seems to be reasonable to believe that the nontrivial
topological structure of the Universe gives indeed a correct explanation to
the DM phenomenon.

\section{The Tully-Fisher relation}

The parameter $\mu $ in (\ref{b4}) has the meaning of the
temperature at which the topology has been tempered (Kirillov \&
Turaev 2002; Kirillov, 2003). We expect that in the very early
Universe it had the order of the Planck temperature $T_{Pl}$
(recall that at $T>T_{Pl}$ quantum gravity effects are thought to
dominate). When, on the further stages of the Universe evolution,
topology changes were suppressed, this parameter evolved as $\mu
\sim T_{Pl}a\left( t_{Pl}\right) /a\left( t\right) $, where $a$ is
the scale factor. Note, however, that the present value of $\mu $
has the sense of the primordial temperature. As it is extremely
small: $\mu \sim 10^{-23}T_{\gamma }$ where $T_{\gamma }$ is the
temperature of CMB radiation ($T_{\gamma }\simeq 2.7K$), we have
to admit the existence of a specific phase in the past when the
nontrivial topological structure might decay (Kirillov 2002,
2003), causing a certain re-heating of matter.
%
During the decay phase, $\mu a$ was a decreasing function of time,
i.e. the scale $R_{0}=\pi /\left( 2\mu \right) $, that corresponds
to the cross-over from the standard Newton's law to the
logarithmic behavior of the potential of a point mass, grew faster
than the scale factor $a(t)$.

Note that the homogeneity of the Universe requires the total mass
distribution (luminous plus dark components) to have a constant density in
space. With the bias of form (\ref{b3}), (\ref{b4}), this corresponds to a
fractal distribution of baryons (Kirillov 2002, 2003), i.e. the number of
baryons within the sphere of a radius $R>R_{0}$ behaves as
\begin{equation}
N_{b}\left( R\right) \simeq \nu R^{D}  \label{frd}
\end{equation}
with $D\simeq 2$, while for $R<$ $R_{0}$ the fractal distribution
is unstable (for the Newton's law restores and baryons dominate
over the DM). The increase of $\ R_{0}/a$ allows one to assume
that in the very early Universe there was a moment $t_{\ast }$
when $N_{b}\left( R_{0}\left( t_{\ast }\right) \right) $ $<1$,
i.e. baryons had the fractal distribution (\ref{frd}) at all
scales. After the topology decay phase, as the scale $ R_{0}\left(
t\right) $ ``jumps'' towards a new, higher value, the fractal
distribution is preserved at scales larger than $R_{0}$, but it
becomes unstable on smaller scales. The instability develops and
baryons under a certain scale of order $R_{0}$ start to
redistribute, governed by Newtonian dynamics. This means that we
can relate $R_{0}$ to the scale of galaxy formation. Then,
according to (\ref{frd}), we should expect the number of baryons
in a galaxy to be
\begin{equation}
N_{b}\simeq \nu ^{\prime }R_{0}^{D},  \label{nbtf}
\end{equation}
with the values of $R_{0}$ and $\nu ^{\prime }$ corresponding to
the moment when a galaxy started to form. Note, however, that
during the formation of a galaxy the value of $ R_{0}$ switches
off from the Hubble expansion (Kirillov 2006), i.e. law
(\ref{nbtf}) remains valid for the present-time values of $N_{b}$
and $R_{0}$.

It is easy to see that this relation leads directly to the
Tully-Fisher law $ L\sim V^{4}$ (Tully \& Fisher 1977), where $L$
is the luminosity and $V$ is the rotation velocity of a galaxy.
Indeed, recall that $\ R_{0}$ fluctuates in space: small spatial
fluctuations of the primordial temperature $\mu $ ($ \Delta \mu
/\mu $ $\sim \Delta T_{\gamma }/T_{\gamma }$) represent seeds for
the present-time scatter in the local value of $\mu =\pi /\left(
2R_{0}\right) $ in different galaxies. Accordingly, the masses of
galaxies $ M\sim m_{b}N_{b}$ (where $m_{b}$ is the baryon mass)
fluctuate as
\begin{equation}
M\simeq m_{b}\nu ^{\prime }R_{0}^{D}.  \label{R}
\end{equation}
This fixes the choice of
\begin{equation}
\mu =\mu _{_{g}}\simeq \frac{\pi }{2}\left( \frac{M_{g}}{m_{b}\nu
^{\prime }}\right) ^{-1/D}  \label{mutf}
\end{equation}%
in the bias (\ref{b3}), (\ref{b4}) for any given galaxy.
Sufficiently far from the center, the galaxy can be considered as
a point-like object, so (\ref{br3}),(\ref{b3}) yield the following
law for the gravitational acceleration at a sufficient distance
from the edge of the optical disc:
\begin{equation}
g=\frac{GM_{g}}{r^{2}}\left( 1+\frac{2}{\pi }\left( \mu _{g}r-\sin \left(
\mu _{_{g}}r\right) \right) \right) .  \label{NL}
\end{equation}%
As $V_{\infty }^{2}/r=g\left( r\right) $, the Tully-Fisher
relation for the asymptotic rotation velocity $V_{\infty }$
follows now from (\ref{NL}), (\ref{mutf}):
\begin{equation}
V_{\infty }^{2}\simeq \frac{2}{\pi }GM_{_{g}}\mu _{_{g}}\qquad
\Longrightarrow \qquad L_{g}\sim M_{g}\simeq \left(
\frac{V_{\infty }}{a} \right) ^{\beta }\ ,  \label{av}
\end{equation}%
with $\beta =\frac{2D}{D-1}$ and $a^{2}=\frac{2}{\pi }G\left( m_{b}\nu
^{\prime }\right) ^{1/D}$. Thus, in our interpretation of the DM, the
Tully-Fisher law reduces to relation (\ref{mutf}) which, in turn, can be
read as an indication of the fractality of the primordial distribution of
baryons with the dimension $D\simeq 2$ (see (\ref{frd})).

\section{Rotation Curve of Spirals}

Let us now compute the rotation curve (RC) of a galaxy modeled by
an infinitely thin disk with surface mass density distribution
$\rho _{L}=\sigma e^{-r/R_{D}}\delta \left( z\right) $. From
(\ref{br1}), (\ref{b4} ), we find for the DM halo density (we use
the notations $x=r/R_{D}$ and $ \lambda =\mu R_{D}$)
\begin{equation}
\rho _{H}\left( \vec{x}\right) =\frac{\lambda \sigma }{2\pi ^{2}R_{D}}\int
\frac{e^{-y}}{\left\vert x-y\right\vert ^{2}}\left( 1-\cos \left( \lambda
\left\vert x-y\right\vert \right) \right) d^{2}y
\end{equation}%
where $y$ lies on the plane $z=0$, while $\vec{x}$ is the 3-dimensional
vector. For the sake of convenience we present the Fourier transform
\begin{equation}
\rho _{H}\left( k,k_{z}\right) =\frac{\mu
}{\sqrt{k^{2}+k_{z}^{2}}}\frac{M_L}{ \left( \left( kR_{D}\right)
^{2}+1\right) ^{\frac{3}{2}}}\theta \left( \mu -
\sqrt{k^{2}+k_{z}^{2}}\right)  \label{dhf}
\end{equation}%
where $\theta $ is the step function: $\theta \left( u\right) =0$ for $u<0$,
and $\theta \left( u\right) =1$ for $u>0$, and $M_{L}=2\pi \sigma R_{D}^{2}$
is the (non-dark) mass of the galaxy.

First of all we note that this distribution is quite consistent with the
observed cored distribution (Gentile et al., 2004). Indeed, in the central
region of the galaxy
\begin{equation}
\rho _{H}\left( 0\right) =\frac{M_{L}}{\left( 2\pi \right) ^{2}R_{D}^{3}}
\lambda \ln \left( 1+\lambda ^{2}\right) ,
\end{equation}
while for $x\gg 1$ we find
\begin{equation}
\rho _{H}\left( \vec{x}\right) \approx \frac{2\rho _{H}\left( 0\right) }{\ln
\left( 1+\lambda ^{2}\right) }\frac{\left( 1-\cos \left( \lambda x\right)
\right) }{x^{2}}.
\end{equation}
If we neglect the oscillating term and compare this with the
pseudo-isothermal halo $\rho =\rho _{0}\frac{\alpha ^{2}}{\alpha ^{2}+x^{2}}$
we find for the core radius
\begin{equation}
\alpha ^{2}=\frac{R_{C}^{2}}{R_{D}^{2}}=\frac{2}{\ln \left( 1+\lambda
^{2}\right) }.  \label{a}
\end{equation}
According to PSS, the core radius can be estimated as
\begin{equation}
\alpha =4.8\left( L/L_{\ast }\right) ^{1/5}
\end{equation}
with $\log L_{\ast }=10.4$, which makes $\lambda $ a certain function of the
luminosity.

Consider now circular velocities predicted by the above mass distributions.
For the disk contribution to the equilibrium circular velocity, we get PSS
\begin{eqnarray}
\frac{V_{D}^{2}}{V_{\infty }^{2}} &=&f_{D}\left( x,\lambda \right) = \\
&=&\frac{\pi }{4}\frac{x^{2}}{\lambda }\left( I_{0}\left( \frac{x}{2}\right)
K_{0}\left( \frac{x}{2}\right) -I_{1}\left( \frac{x}{2}\right) K_{1}\left(
\frac{x}{2}\right) \right) ,  \nonumber
\end{eqnarray}%
and for the Dark Halo contribution we find from (\ref{dhf}) the expression
\begin{equation}
\frac{V_{H}^{2}}{V_{\infty }^{2}}=f_{H}\left( x,\lambda \right)
=\frac{x}{ \lambda }\int_{0}^{\lambda }\frac{\sqrt{\lambda
^{2}-k^{2}}}{\left( \sqrt{ \left( k^{2}+1\right) }\right)
^{3}}J_{1}\left( kx\right) dk,
\end{equation}%
where $J_{n}$, $I_{n}$, $K_{n}$ are the Bessel and the modified Bessel
functions and $V_{\infty }^{2}=\frac{GM_{g}}{R_{D}}\frac{2}{\pi }\lambda $.
Thus for the rotation curve we find the expression
\begin{equation}
V^{2}\left( x,\lambda \right) =V_{\infty }^{2}\left( f_{D}\left( x,\lambda
\right) +f_{H}\left( x,\lambda \right) \right) .  \label{URC}
\end{equation}
As we see, the shape of the rotation curve depends indeed on one
parameter $ \lambda $. Via relation (\ref{R}), or equivalently
(\ref{av}), $\lambda $ is expressed as a function of the total
number of baryons $N_{b}$ in the galaxy; there is, however, an
uncertainty in $\lambda $ due to the variation of the ratio $M/L$
for different galaxies. At the moment of the galaxy formation
$R_{D}\sim R_{0}$ which corresponds to the same initial value $
\lambda \sim 1$ in all galaxies. On subsequent stages of the
evolution $ \lambda $ becomes different in different objects.
Indeed in smaller galaxies supernovae are more efficient in
removing the gas from the central (stellar forming) region of a
galaxy than in bigger galaxies (e.g., see Shankar et al., 2006 and
references therein) and this creates the fact that in smaller
objects the disc has a smaller baryonic density (a lower surface
brightness) and the ratio $\lambda \simeq R_{D}/R_{0}\gg 1$.

To compare expression (\ref{URC}) with that from PSS we rewrite it as
\begin{equation}
\frac{V^{2}\left( x,\lambda \right) }{V_{opt}^{2}}=\frac{f_{D}\left(
x,\lambda \right) +f_{H}\left( x,\lambda \right) }{f_{D}\left( 3.2,\lambda
\right) +f_{H}\left( 3.2,\lambda \right) }  \label{urc1}
\end{equation}
where $x=3.2$ corresponds to the optical radius of a galaxy. The plot of
this curve for different values of $\lambda $ is presented in Fig.1.

\begin{figure*}
\centerline{\psfig{figure=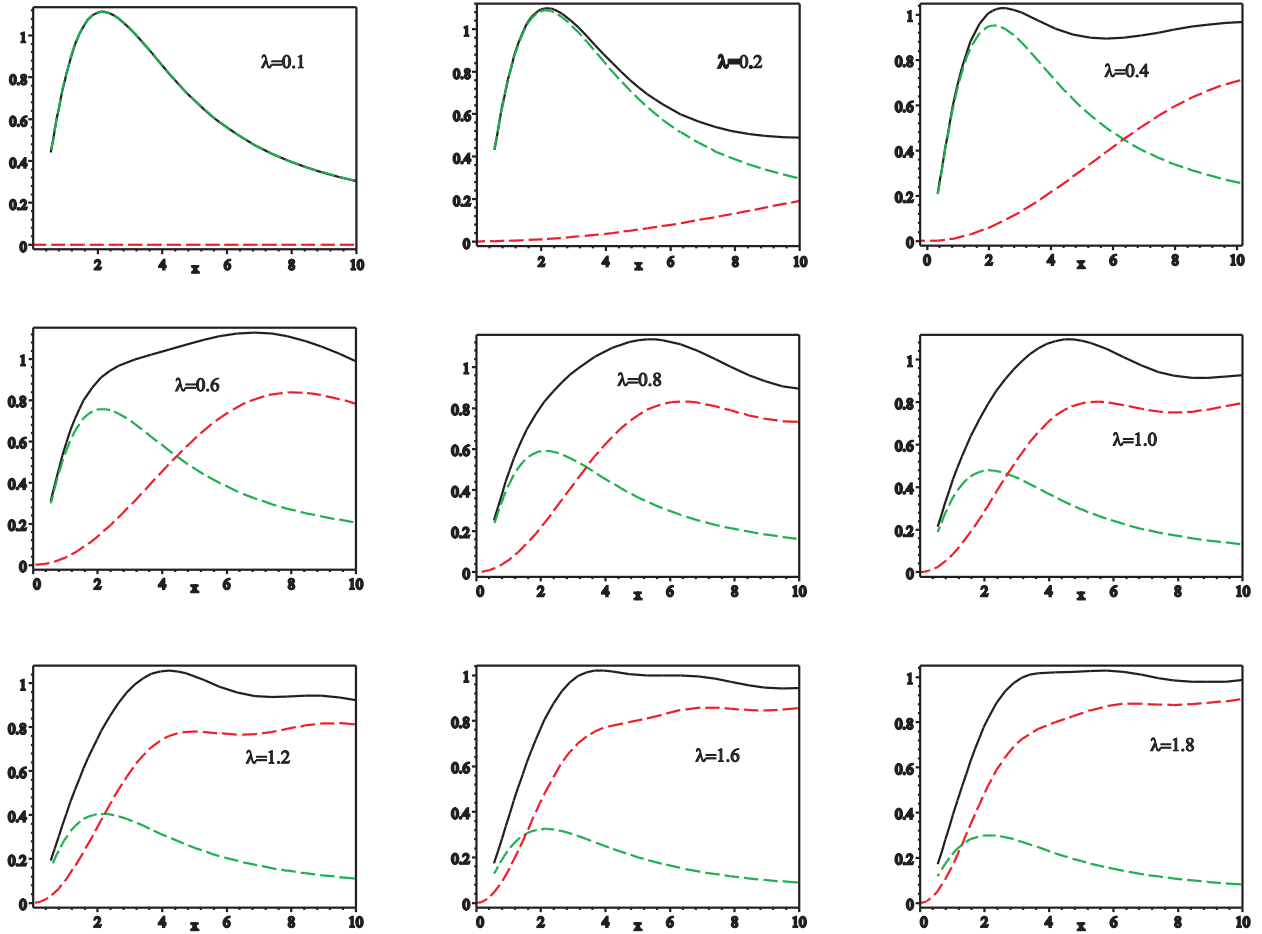,width=17.7cm}} \caption{The
rotation curves $V^{2}\left( x\right) /V_{opt}^{2}$ vs $x=r/R_{D}$
for different values of $\protect\lambda $ . The green and red
dashed lines give the visible and DM contributions respectively,
while the black line gives the sum. We see that with the decrease
of the luminosity (increase of $\protect\lambda $) DM fraction
increases in agreement with PSS. } \label{curves}
\end{figure*}

While the similarity of our RC (\ref{urc1}) with the empirical URC
of PSS is quite good, we note that the topological bias (\ref{b4})
predicts a new feature in Rotation Curves -- specific oscillations
in the DM density with the characteristic wavelength $\ell =2\pi
/\lambda $ (or in dimensional units $\ell \sim M_{g}^{1/D}$).
Indeed at a sufficient distance from the edge of the optical disc
(i.e., as $x\gg 3.2$) a galaxy can be considered as a point-like
object. Then from (\ref{NL}) for the rotation
velocity we find the expression%
\begin{equation}
\frac{V^{2}\left( x\right) }{V_{\infty }^{2}}=\frac{\pi }{2\lambda
x}+1- \frac{\sin \left( \lambda x\right) }{\lambda x}
\label{urc2}
\end{equation}%
which shows the presence of a specific oscillations with the
decaying amplitude $1/(\lambda x)$. Such oscillations are, in
turn, rather difficult (though possible) to extract from
observations. Indeed in the case of HSB (high surface brightness)
galaxies when $\lambda \lesssim 1$ (i.e., for rather long periods
$\ell \gtrsim 2\pi $) the expression (\ref{urc2}) gives a very
good quantitative approximation to the exact formula (\ref{urc1})
starting already from $x=x_{opt}=3.2$. However the reliable RC
data available extend usually not more than to $x=(2-3)x_{opt}$.
In this range oscillations are not established yet and the
beginning of oscillations is seen (e.g., see Fig.1.) as not flat
RC slopes. The slopes observed are known to take values between
$0.2$ and $-0.2$ (e.g., see PSS). In the case of LSB galaxies
$\lambda \gg 1$ ($\ell \ll 2\pi $), the amplitude of oscillations
is somewhat suppressed $\sim 1/\lambda $ and the small amount of
the stellar mass in the range $x_{opt}<x<3x_{opt}$ considerably
smooths such oscillations which results in a some deviation of
(\ref{urc2}) from the exact expression (\ref{urc1}). Moreover, in
deriving (\ref{urc1}) we do not take into account the presence of
gas which due to supernovae does not trace the brightness, i.e.,
it deviates the exponential profile. Essentially this is true for
LSB galaxies. Thus, to observe such oscillations we have either to
measure velocities for sufficiently large distances $\sim
10x_{opt} $ (e.g., for HSB), or to improve the accuracy of
available observational data in LSB.

\section{Discussion and Conclusions}

As it can be seen from Fig.1, the topological bias (\ref{b3})
predicted in (Kirillov \& Turaev 2002; Kirillov 2003) shows quite
a good agreement with observations. Indeed, it repeats all
features of the empirical URC of PSS: the amount of DM
progressively increases with the decrease of the luminosity (cf.
PSS), DM shows the cored distribution (cf. Gentile et al., 2004)
with the strong correlation (\ref{a}) between the core radius and
the disk size (cf. Donato, et al., 2004), and the Tully-Fisher
relation (Tully \& Fisher 1977) is explicitly present (see
(\ref{av})). There is no doubt that the tuning of a single free
parameter $M/L$ allows to fit any RC. At least, it is claimed in
(e.g., see Milgrom \& Sanders 2005 and references therein) for the
RCs obtained via Milgrom algorithm MOND (Milgrom  1983), and our
RCs are phenomenologically quite close to those, although the
physics in our approach is completely different. In this respect
we can claim that the topological bias gives a rigorous basis for
applying the MOND-type algorithm in galaxies (which however allows
for the Tully-Fisher relation to have $\beta \neq 4$). Therefore,
there is enough evidence that bias (\ref{br1}), (\ref{b4}) gives
an adequate description of galaxies.

We repeat that our approach produces as good fit to the observed
RCs as the Milgrom algorithm (known to be quite successful
empirically (Milgrom \& Sanders 2005, see however Gentile et al.
2004, Donato et al. 2004)) can do. However, contrary to MOND, our
theory remains linear in weak fields, and the superposition of
forces holds. In fact, our approach does not presume any
modification of the theory and basic equations: there is actually
no modification of gravity, while the bias appears merely as a
result of a disagreement between the actual topology of the
physical space and that of the flat space (e.g., see Sec. 2 in
Kirillov 2006). Thus, there are all reasons to believe that the DM
phenomenon has indeed the topological origin.

Once we accept the topological origin of the bias (\ref{br1}), (\ref{b4}),
the Tully-Fisher relation
\begin{equation}
L\sim V_{\infty }^{\beta }
\end{equation}%
with $\beta =\frac{2D}{D-1}\simeq 4$ serves as a strong indication
of the fractal behavior in the primordial distribution of baryons
with the dimension $D\simeq 2$. Such fractal distribution changes
essentially the estimate for the baryon number density in the
Universe (e.g., see Kirillov 2006). The currently accepted
post-WMAP cosmology has (roughly): $\Omega _{total}=1$, $\Omega
_{\Lambda }\sim 0.7$, $\Omega _{DM}\sim 0.25$, and $ \Omega
_{b}\sim 0.05$, which implies $\Omega _{DM}/\Omega _{b}\sim 5$. We
stress that such estimates are model dependent, for they are
strongly based on the standard model (e.g., the content,
evolution, the homogeneity of the baryon distribution, the power
law of initial spectrum of perturbations, etc.)\footnote{ For
linear perturbations there are no doubts that if we take the
observed spectrum of $\Delta T/T$ and fix an arbitrary model $a(
t)$, we find in a unique way the initial primordial spectrum of
$\Delta \rho /\rho $. To fix the model we have to know both the
measured $\Delta T/T$ and the initial values of $\Delta \rho /\rho
$ which a priori are unknown.}. Moreover, the direct count of the
number of baryons gives $ \Omega _{b}\sim 0.003$ for the whole
nearby Universe out to the radius $\sim 300h_{50}^{-1}Mpc$ (e.g.,
see Persic \& Salucci, 1992), which means that in the standard
cosmological models most of baryons are somewhere hidden.

When the topological bias is accepted such estimates require
essential revision. Indeed, according to (\ref{b3}) the
topological bias modifies the Newton's law in the range of scales
$\mu >k>\kappa $ where the equilibrium distribution of baryons has
the fractal behavior. On scales $ k<\kappa $ the standard Newton's
law restores (and baryons cross over to the homogeneity) but
dynamically every particle becomes heavier in $1+\mu /\kappa $
times, which gives for the effective DM fraction $\Omega
_{DM}/\Omega _{b}\sim $ $1+\mu /\kappa $. The scale $ R_{0}=\pi
/\left( 2\mu \right) $ is directly measured in galaxies by RCs and
is estimated as a few Kpc. However the mean value $<R_{0}>$ for
the homogeneous Universe should be $10^{2}$ times bigger (Kirillov
2006). The maximal scale $1/\kappa $ is the scale where the
primordial fractal distribution of baryons crosses over to the
homogeneity. This scale is not so easy to measure without a
detailed investigation. Indeed, the large-scale structure, e.g.,
the existence of huge ($\sim 100-200Mpc$) voids with no galaxies
inside and thin ($\sim 1-5Mpc$) walls filled with galaxies, fits
quite well into the fractal picture and suggests only the lower
boundary $1/\kappa >100-200Mpc$. This gives for the DM fraction
$\Omega _{DM}/\Omega _{b}>10^{2}-10^{3}$ which is consistent with
the observed value $\Omega _{b}\sim 0.003$. However the maximal
possible value $1/\kappa \sim R_{H}$ ($R_{H}$ is the Hubble
radius) which gives $\Omega _{DM}/\Omega _{b}\sim 10^{5}$ cannot
be excluded. To avoid misunderstanding we stress that the
topological nature of the bias makes the fractal distribution to
be equilibrium and consistent with the homogeneity of the metric
and the observed CMB fluctuations $\Delta T/T$ (e.g., see Kirillov
2006). The topological nature means that the same bias appears in
all interactions. If the bias would not modify the electromagnetic
field, then the fractal distribution of baryons would be in a
severe conflict with observations and surely had to be rejected as
it does take place in the standard models (e.g., the fractal
distribution produces too strong fluctuations $\Delta T/T$ $\sim
\Delta \rho _{b}/\rho _{b}\sim \mu /\kappa $ ). The topological
nature of the bias however creates the fact that the Coulomb force
and all Green functions are also modified at galaxy scales
(Kirillov \& Turaev 2002, Kirillov 2006) which reduces $\Delta
T/T$ to the observed value $\Delta \rho _{total}/\rho _{total}$
$\sim  \kappa /\mu$ $\sim 10^{-5}$ (e.g., for sufficiently remote
objects $<\mu >r\gg 1 $ the apparent luminosity has to behave as
$\ell \sim L/r^{D-1}$ instead of $1/r^{2}$ which gives for the
number of objects brighter than $\ell $ somewhat higher (with
respect to $D=3$) value $N\left( \ell \right) \approx \nu
r^{D}\left( \ell \right) \approx 1/\ell ^{D/( D-1) }$). Thus the
topological bias and the observational definition of $1/\kappa $
requires the careful and thorough revision of the standard model
and all basic formulas.

In conclusion, we point out that bias (\ref{b4}) predicts the existence of
specific oscillations in the distribution of DM with the characteristic
wavelength $\sim M_{g}^{1/D}\sim \sqrt{L}$. When the observational data
allow, this can be used to verify the theory and, thus, to make a more
definite conclusion on the nature of DM.

\section{Acknowledgement}
We acknowledge P. Salucci for valuable comments and the referee's
valuable advice which helped us to improve essentially the
presentation of this work. This research was supported in part by
the Center for Advanced Studies in the Ben-Gurion University of
the Negev.


\begin{thebibliography}{\citeauthoryear{Weldrake, de Blok \&
Walter}{Weldrake et al.}{2003}}
\bibitem[\protect\citeauthoryear{Borriello et al. }{2003}]{CE2} 
Borriello, A., Salucci, P., Danese, L., 2003, MNRAS, 341, 1109.

\bibitem[\protect\citeauthoryear{de Blok \& Bosma} {2002}]{Core3} de Blok, W. J. G.,
Bosma, A., 2002, A\&A, 385, 816.

\bibitem[\protect\citeauthoryear{Donato, Gentile, \& Salucci }{2004}]{S04}
Donato, F., Gentile, G., and Salucci, P., 2004, MNRAS, 353, L17.

\bibitem[\protect\citeauthoryear{Gentile et al.}{2004}]{Core} Gentile, G., Salucci,
P., Klein, U., Vergani, D., Kalberla, P., 2004, MNRAS, 351, 903.

\bibitem[\protect\citeauthoryear{Gerhard et al. }{2001}]{CE} 
Gerhard, O., Kronawitter, A., Saglia, R.P., Bender, R., 2001, ApJ,
121, 1936.

\bibitem[\protect\citeauthoryear{Kirillov }{1999}]{K99}
Kirillov, A.A., 1999, JETP, 88, 1051.

\bibitem[\protect\citeauthoryear{Kirillov \& Turaev }{2002}]{KT02} 
Kirillov, A.A., Turaev, D., 2002 Phys. Lett. B, 532,185.

\bibitem[\protect\citeauthoryear{Kirillov }{2002}]{k02} 
Kirillov, A.A., 2002, Phys. Lett. B, 535, 22.

\bibitem[\protect\citeauthoryear{Kirillov }{2003}]{K03} 
Kirillov, A.A., 2003, Phys. Lett. B, 555, 13.

\bibitem[\protect\citeauthoryear{Kirillov }{2006}]{K06} 
Kirillov, A.A., 2006, Phys. Lett. B, 632, 453.

\bibitem[\protect\citeauthoryear{Labini et al.}{1998}]{Fra3} Labini, S.F., Montuori,
M., Pietronero, L., 1998 Phys. Rep. 293, 66.

\bibitem[\protect\citeauthoryear{Milgrom }{1983}]{mil} Milgrom, M., 1983, ApJ, 270,
365

\bibitem[\protect\citeauthoryear{Milgrom \& Sanders}{2005}]{ms} Milgrom, M.,
Sanders, R.H., 2005, MNRAS, 356, 45.



\bibitem[\protect\citeauthoryear{Navarro et al. }{1996}]{NFW} Navarro, J.F., Frenk,
C.S., White, S.D.M., 1996, ApJ, 462, 563


\bibitem[\protect\citeauthoryear{Persic, Salucci \& Stel} {1996}]{PSS} Persic, M.,
Salucci, P., Stel, F., 1996, MNRAS, 281, 27.

\bibitem[\protect\citeauthoryear{Persic, Salucci \& Stel} {1996}]{PS} Persic, M.,
Salucci, P., 1992, MNRAS, 258, 14.

\bibitem[\protect\citeauthoryear{Pietronero }{1987}]{Fra} Pietronero L., 1987,
Physica, A144, 257.

\bibitem[\protect\citeauthoryear{Ruffini et al.}{1988}]{Fra2} Ruffini R., Song D.J.,
and Taraglio S., 1988 A\&A, 190, 1.

\bibitem[\protect\citeauthoryear{Shankar et al.}{2006}]{SL} Shankar, F.; Lapi, A.;
Salucci, P.; De Zotti, G.; Danese, L., 2006, ApJ, 643, 14S.

\bibitem[\protect\citeauthoryear{Tully \& Fisher }{1977}]{TF} 
Tully, R.B., Fisher, J.R. 1977, A\&A, 54, 661.


\bibitem[\protect\citeauthoryear{Weldrake, de Blok \& Walter}{Weldrake et al.}{2003}]%
{Core2} Weldrake, D.T.F., de Blok, W. J. G., Walter, F., 2003,
MNRAS, 340, 12.
\end{thebibliography}
\end{document}